\documentclass[12pt]{iopart}
\usepackage{iopams}
\usepackage{setstack}
\usepackage[dvips]{graphicx}
\usepackage{epstopdf}
\usepackage{subfigure}
\usepackage{color}

\begin{document}
\title{Spin filtering in a Rashba-Dresselhaus-Aharonov-Bohm double-dot interferometer}
\author{Shlomi Matityahu$^1$, Amnon Aharony$^1$$^,$$^2$$^,$$^3$, Ora Entin-Wohlman$^1$$^,$$^2$$^,$$^3$ and Seigo Tarucha$^4$$^,$$^5$}
\address{$^1$ Department of Physics, Ben-Gurion University, Beer Sheva 84105, Israel}
\address{$^2$ Ilse Katz Center for Meso- and Nano-Scale Science and Technology, Ben-Gurion University, Beer Sheva 84105, Israel}
\address{$^3$ Raymond and Beverly Sackler School of Physics and Astronomy, Tel Aviv University, Tel Aviv 69978, Israel}
\address{$^4$ Department of Applied Physics, University of Tokyo, Bunkyo-ku, Tokyo, 113-8656, Japan}
\address{$^5$ ICORP (International Cooperative Research Project) Quantum Spin Information Project, Atsugi-shi, Kanagawa, 243-0198, Japan}
\ead{aaharony@bgu.ac.il}
\begin{abstract}
We study the spin-dependent transport of spin-1/2 electrons
through an interferometer made of two elongated quantum dots or
quantum nanowires, which are subject to both an Aharonov-Bohm flux
and (Rashba and Dresselhaus) spin-orbit interactions. Similar to
the diamond interferometer proposed in our previous papers [Phys.
Rev. B {\bf 84}, 035323 (2011); Phys. Rev. B {\bf 87}, 205438
(2013)], we show that the double-dot interferometer can serve as a
perfect spin filter due to a spin interference effect. By
appropriately tuning the external electric and magnetic fields
which determine the Aharonov-Casher and Aharonov-Bohm phases, and
with some relations between the various hopping amplitudes and
site energies, the interferometer blocks electrons with a specific
spin polarization, independent of their energy. The blocked
polarization and the polarization of the outgoing electrons is
controlled solely by the external electric and magnetic fields and
do not depend on the energy of the electrons. Furthermore, the
spin filtering conditions become simpler in the linear-response
regime, in which the electrons have a fixed energy. Unlike the
diamond interferometer, spin filtering in the double-dot
interferometer does not require high symmetry between the hopping
amplitudes and site energies of the two branches of the
interferometer and thus may be more appealing from an experimental
point of view.
\end{abstract}
\maketitle
\section{Introduction}
\label{Intro} Spin-dependent electrons transport in
low-dimensional mesoscopic systems has recently drawn much
attention due to its potential for future electronic device
applications in the field of spintronics
\cite{PGA98,WSA01,ZI04,BSD10}. This new emerging field deals with
the active manipulation of the electron's spin (and not only its
charge). Adding the spin degree of freedom to the conventional
charge-based technology has the potential advantages of
multifunctionality, longer decoherence times and lengths,
increased data processing speed, decreased electric power
consumption, and increased integration densities compared with
conventional semiconductor devices \cite{PGA98,WSA01}. In addition
to the improvement of contemporary technology, spintronics may
also contribute to the field of quantum computation and quantum
information, in which the quantum information may be contained in
the unit vector along which the spin is polarized
\cite{Nielsen&Chuang}. Writing and reading information on a spin
qubit are thus equivalent to polarizing the spin along a specific
direction and identifying the direction along which the spin is
polarized, respectively. Hence, a major aim of spintronics is to
build mesoscopic spin valves (or spin filters), which generate a
tunable spin-polarized current out of unpolarized electron
sources. Spin filters can also be used as spin analyzers, which
read this information by identifying the polarization directions
of incoming polarized beams. A priori, a straightforward way to
realize such devices is by using ferromagnets that inject and/or
collect polarized electrons \cite{JBT07}. However, connecting
ferromagnets to semiconductors is inefficient, due to a large
impedance mismatch between them \cite{ZI04,SG00,SG02,SG05,TT11}.
Therefore, efforts are being made to model and fabricate
spintronic devices using intrinsic properties of mesoscopic
systems such as strong spin-orbit interaction (SOI). This paradigm
of spintronics involves small fields without the need for
ferromagnetism at all \cite{KY04,ADD09}.

In a two-dimensional electron gas (2DEG), formed in mesoscopic
structures made of narrow-gap semiconductor heterostructures, the
SOI has the general form $\mathcal{H}^{}_{\rm{SO}}=(\hbar
k^{}_{\rm{SO}}/m)(\bpi\cdot\bsigma)$ \cite{Winkler}. Here
$k^{}_{\rm{SO}}$ characterizes the SOI strength, $\bpi$ is a
linear combination of the electron momentum components $p^{}_{x}$
and $p^{}_{y}$ and $m$ is the effective mass. The vector of Pauli
matrices $\bsigma$ is related to the electron spin via
$\bi{S}=\hbar\bsigma/2$. We distinguish between two special cases
of the linear (in the momentum) SOI, namely the Rashba SOI
\cite{Rashba1,Rashba2} and the Dresselhaus SOI \cite{Dresselhaus}.
The Rashba SOI is present in narrow-gap semiconductor
heterostructures with a confining potential well which is
asymmetric under space inversion. For an electric field
perpendicular to the interferometer plane (defined as the $z$
axis), this SOI has the form
\begin{eqnarray}
\label{Rashba Hamiltonian} \mathcal{H}^{}_{\rm{R}}=\frac{\hbar
k^{}_{\rm{R}}}{m}\left(p^{}_{y}\sigma^{}_{x}-p^{}_{x}\sigma^{}_{y}\right).
\end{eqnarray}
The coefficient $k^{}_{\rm{R}}$ depends on the magnitude of the
electric field and can be controlled by a gate voltage, as shown
in several experiments \cite{NJ97,HJP98,GD00,KT02,KM06,BT06,LD12}.
The Dresselhaus SOI is a consequence of a host crystal which lacks
bulk inversion symmetry. For a 2DEG the linear Dresselhaus SOI is
given by
\begin{eqnarray}
\label{Dresselhaus Hamiltonian}
\mathcal{H}^{}_{\rm{D}}=\frac{\hbar
k^{}_{\rm{D}}}{m}\left(p^{}_{x}\sigma^{}_{x}-p^{}_{y}\sigma^{}_{y}\right),
\end{eqnarray}
where $k^{}_{\rm{D}}$ is a material constant which is proportional
to $1/d^{2}$, with $d$ the quantum well thickness \cite{LJ90}. It
depends weakly (if at all) on the external field. These SOIs can
be interpreted as a Zeeman interaction in a momentum-dependent
effective magnetic field. As the electron propagates in the
presence of these SOIs, its spin precesses around this effective
magnetic field. As a consequence, after propagating a distance $L$
in the direction of the unit vector $\hat{\bi{g}}$, the electron's
spinor $|\chi\rangle$ transforms into
$|\chi'\rangle=U|\chi\rangle$ with the SU(2) matrix
$U=e^{i\bi{K}\cdot\bsigma}$ \cite{OY92,EO05}. Here, the vector
$\bi{K}$ is
\begin{eqnarray}
\label{Rotation matrix1}
\bi{K}=\alpha^{}_{\rm{R}}\left(-g^{}_{y},g^{}_{x},0\right)+\alpha^{}_{\rm{D}}\left(-g^{}_{x},g^{}_{y},0\right),
\end{eqnarray}
with the dimensionless coefficients
$\alpha^{}_{\rm{R},\rm{D}}\equiv k^{}_{\rm{R},\rm{D}}L$. The
SOI-related phase of the unitary matrix $U$ is known as the
Aharonov-Casher (AC) phase \cite{AY84}. Below we use the unitary
matrix $U$ and the parameters $\alpha^{}_{\rm{R},\rm{D}}$ to
characterize the hopping between adjacent bonds in the presence of
SOI.

Recently, several groups proposed spin filters based on a single
loop, subject to both electric and magnetic fields perpendicular
to the plane of the loop \cite{CR06,HN07,AA11}. The phases of
these waves include the AC phase and the Aharonov-Bohm (AB) phase
\cite{AY59}, which results from a magnetic flux $\Phi$ penetrating
the loop. When an electron goes around such a loop, its wave
function gains an AB phase $\phi=2\pi\Phi/\Phi^{}_{0}$, where
$\Phi^{}_{0}=hc/e$ is the flux quantum ($c$ is the speed of light
and $e$ is the electron charge). The combined effect of the SOI
and the AB flux is to transform the spinor $|\chi\rangle$ of an
electron that goes around a loop into
$|\chi'\rangle=u|\chi\rangle$, where the unitary matrix $u$ is of
the form
\begin{eqnarray}
\label{Rotation matrix2}
u=u^{}_{\rm{AB}}u^{}_{\rm{SOI}}=e^{-i\phi+i\bomega\cdot\bsigma}=e^{-i\phi}\left(\cos\omega+i\sin\omega\:\hat{\bomega}\cdot\bsigma\right).
\end{eqnarray}
Here $u^{}_{\rm{AB}}=e^{-i\phi}\bi{I}$ ($\bi{I}$ is the $2\times
2$ unit matrix) is the diagonal transformation matrix due to the
AB flux and $u^{}_{\rm{SOI}}=e^{i\bomega\cdot\bsigma}$ is the
transformation matrix due to the SOI. The latter is a product of
matrices of the form $e^{i\bi{K}\cdot\bsigma}$ discussed above,
each coming from the local SOI on a segment of the loop
\cite{AA11}. We neglect the Zeeman term
$\mathcal{H}^{}_{\rm{Z}}=\left(g\mu^{}_{\rm{B}}/2\right)\bsigma\cdot\bi{B}$
($g$ is the Land\'{e} factor and $\mu^{}_{\rm{B}}$ is the Bohr
magneton). Even though $g$ can be large in low-dimensional
systems, the Zeeman term is much smaller than the SOI terms
(\ref{Rashba Hamiltonian}) and (\ref{Dresselhaus Hamiltonian}) in
the magnetic fields considered.

In a previous paper we proposed a diamond interferometer which
combines the AB and AC phases to filter a specific spin direction
\cite{AA11}. The filtered direction can be tuned by the external
electric and magnetic fields. Moreover, the transmission of the
outgoing spin-polarized electrons can be tuned to unity in a wide
range of energies. Recently, we generalized this interferometer by
including a possible leakage of electrons out of the
interferometer \cite{MS13}. We have shown that spin filtering is
still possible in a non-unitary transport, even though the
transmission is inherently less than unity.

A major advantage of the diamond interferometer is that full spin
filtering can be achieved independent of the electron energy.
However, the conditions for full spin filtering, independent of
the electron energy, require perfect symmetry between the two
branches of the interferometer \cite{AA11,MS13}. Having fulfilled
these symmetry relations, the polarization of the outgoing
electrons is independent of energy and completely determined by
the AB and AC phases \cite{AA11,MS13} (and therefore by the
external electric and magnetic fields perpendicular to the
interferometer plane). Unfortunately, realizing a highly symmetric
interferometer in experiments may be a difficult task. It is thus
desirable to have a perfect spin filtering, independent of the
electron energy, in an asymmetric interferometer. As we argue in
this paper, this can be achieved by enlarging the number of
interferometer parameters (such as hopping and site energies).
Below we examine a double-dot interferometer which allows a wide
freedom for the various parameters, thereby simplifies the
experimental realization. The relations between the various
parameters are further simplified if one assumes linear-response
regime (namely, low temperatures and bias voltages), in which
electron transport occurs at a single energy.

\begin{figure}[ht]
\centering
\includegraphics[width=0.5\textwidth]{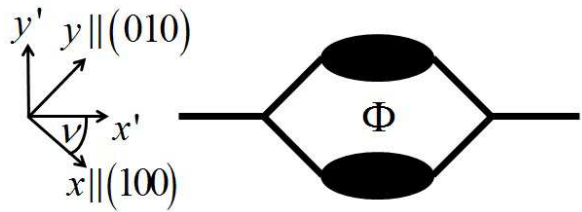}
\caption{\label{fig:square interferometer1} The double-dot
interferometer. The interferometer is penetrated by a magnetic
flux $\Phi$, and its horizontal edges (of length $L$, shown by the
dark ellipses in the figure) are subject to spin-orbit
interactions. The $x$ and $y$ axes are parallel to the
crystallographic $(100)$ and $(010)$ axes.}
\end{figure}

The double-dot interferometer is sketched schematically in figure
\ref{fig:square interferometer1}. It consists of two elongated
quantum dots (QDs) or quantum nanowires (QNs) which are subject to
SOI, and the area of the interferometer is penetrated by an AB
flux. Recently, the electrical control of SOI was demonstrated in
such InAs self-assembled elongated QDs and nanowires
\cite{LD12,TS09,KY11,FC07,NPS10}. The wires which connect the
QDs/QNs are free of SOI. Using scattering theory in the framework
of the tight-binding formalism, we calculate the spin-dependent
transmission through this interferometer. We employ a
one-dimensional tight-binding model, assuming that the
interferometer is composed of quasi one-dimensional wires
\cite{AA11,MS13}.

The paper is organized as follows: in Sec. \ref{Sec 1} we first
define the tight-binding model which we use to study electron
transport in the double-dot interferometer and solve for the
transmission of an arbitrary interferometer (Sec. \ref{Sec 1A}).
Then we find the general conditions for full filtering (Sec.
\ref{Sec 1B}). Finally, we find specific criteria for spin
filtering in the presence of Rashba and Dresselhaus SOIs (Sec.
\ref{Sec 1C}). The results are discussed and summarized in Sec.
\ref{Summary}.
\section{Double-dot interferometer}
\label{Sec 1}
\subsection{Tight-binding model for the double-dot interferometer}
\label{Sec 1A} To study the scattering of a spin-1/2 electron by
the double-dot interferometer with arbitrary SOI and AB flux
(figure \ref{fig:square interferometer1}), we model it as a square
interferometer as shown in figure \ref{fig:square
interferometer2}. Each QD/QN is replaced by a bond connecting two
sites ($a$, $b$ and $c$, $d$ in figure \ref{fig:square
interferometer2}) with the corresponding bonds subject to SOI. We
emphasize that the solution is not limited to this model. One can
model each QD/QN by an arbitrary number of sites $M$. This changes
only the transmission of the spin-polarized electrons, with no
significant effect on the spin filtering conditions and the
polarization direction of the outgoing electrons \cite{MS13}.
\begin{figure}[ht]
\centering
\includegraphics[width=0.5\textwidth]{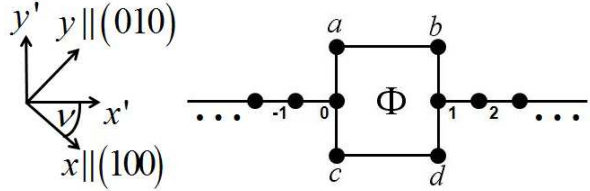}
\caption{\label{fig:square interferometer2} Tight-binding model of
the double-dot interferometer. The bonds $ab$ and $cd$ (of length
$L$) are subject to spin-orbit interactions.}
\end{figure}

In the framework of the nearest neighbors tight-binding model, the
Schr\"{o}dinger equation for the spinor $|\psi^{}_{v}\rangle$ at
site $v$ is written as
\begin{eqnarray}
\label{Tight binding}
\left(\varepsilon-\varepsilon^{}_{v}\right)|\psi^{}_{v}\rangle=-\sum^{}_{u}J^{}_{uv}U^{}_{uv}|\psi^{}_{u}\rangle,
\end{eqnarray}
where $\varepsilon^{}_{v}$ is the site energy, $J^{}_{uv}$ is a
real hopping amplitude and $U^{}_{uv}$ is a $2\times 2$ unitary
matrix which describes the AB and AC phases acquired by an
electron moving from site $u$ to site $v$. The sum in equation
(\ref{Tight binding}) is over the nearest neighbors $u$ of $v$. At
this stage we do not specify the details of these matrices and
hopping amplitudes. Except for the sites $0$, $a$, $b$, $c$, $d$
and $1$ (figure \ref{fig:square interferometer2}), the hopping
amplitude along the leads is $j$ and the site energies on the
leads are set to zero. The leads are free of SOI. Thus, the
dispersion relation of the leads is
$\varepsilon=-2j\cos\left(ka\right)$ with $a$ the lattice
constant. The tight-binding Schr\"{o}dinger equations for the
spinors at sites $0$, $a$, $b$, $c$, $d$ and $1$ are
\begin{eqnarray}
\label{TB}
\left(\varepsilon-\varepsilon^{}_{0}\right)|\psi^{}_{0}\rangle=-J^{}_{0a}U^{\dag}_{0a}|\psi^{}_{a}\rangle-J^{}_{0c}U^{\dag}_{0c}|\psi^{}_{c}\rangle-j|\psi^{}_{-1}\rangle,\nonumber\\
\left(\varepsilon-\varepsilon^{}_{a}\right)|\psi^{}_{a}\rangle=-J^{}_{ab}U^{\dag}_{ab}|\psi^{}_{b}\rangle-J^{}_{0a}U^{}_{0a}|\psi^{}_{0}\rangle,\nonumber\\
\left(\varepsilon-\varepsilon^{}_{b}\right)|\psi^{}_{b}\rangle=-J^{}_{b1}U^{\dag}_{b1}|\psi^{}_{1}\rangle-J^{}_{ab}U^{}_{ab}|\psi^{}_{a}\rangle,\nonumber\\
\left(\varepsilon-\varepsilon^{}_{c}\right)|\psi^{}_{c}\rangle=-J^{}_{cd}U^{\dag}_{cd}|\psi^{}_{d}\rangle-J^{}_{0c}U^{}_{0c}|\psi^{}_{0}\rangle,\nonumber\\
\left(\varepsilon-\varepsilon^{}_{d}\right)|\psi^{}_{d}\rangle=-J^{}_{d1}U^{\dag}_{d1}|\psi^{}_{1}\rangle-J^{}_{cd}U^{}_{cd}|\psi^{}_{c}\rangle,\nonumber\\
\left(\varepsilon-\varepsilon^{}_{1}\right)|\psi^{}_{1}\rangle=-j|\psi^{}_{2}\rangle-J^{}_{b1}U^{}_{b1}|\psi^{}_{b}\rangle-J^{}_{d1}U^{}_{d1}|\psi^{}_{d}\rangle.
\end{eqnarray}
Eliminating $|\psi^{}_{a}\rangle$, $|\psi^{}_{b}\rangle$,
$|\psi^{}_{c}\rangle$ and $|\psi^{}_{d}\rangle$ from equations
(\ref{TB}) one ends up with the equations
\begin{eqnarray}
\label{Effective TB}
\left(\varepsilon-y^{}_{0}\right)|\psi^{}_{0}\rangle=\bi{W}^{\dag}|\psi^{}_{1}\rangle-j|\psi^{}_{-1}\rangle,\nonumber\\
\left(\varepsilon-y^{}_{1}\right)|\psi^{}_{1}\rangle=-j|\psi^{}_{2}\rangle+\bi{W}|\psi^{}_{0}\rangle,
\end{eqnarray}
where
\begin{eqnarray}
\label{y}
y^{}_{0}=\varepsilon^{}_{0}+\frac{J^{2}_{0a}}{\varepsilon-\varepsilon^{}_{a}-\frac{J^{2}_{ab}}{\varepsilon-\varepsilon^{}_{b}}}+\frac{J^{2}_{0c}}{\varepsilon-\varepsilon^{}_{c}-\frac{J^{2}_{cd}}{\varepsilon-\varepsilon^{}_{d}}},\nonumber\\
y^{}_{1}=\varepsilon^{}_{1}+\frac{J^{2}_{b1}}{\varepsilon-\varepsilon^{}_{b}-\frac{J^{2}_{ab}}{\varepsilon-\varepsilon^{}_{a}}}+\frac{J^{2}_{d1}}{\varepsilon-\varepsilon^{}_{d}-\frac{J^{2}_{cd}}{\varepsilon-\varepsilon^{}_{c}}},
\end{eqnarray}
and
\begin{eqnarray}
\label{W}
\bi{W}&=\frac{J^{}_{0a}J^{}_{ab}J^{}_{b1}}{J^{2}_{ab}-\left(\varepsilon-\varepsilon^{}_{a}\right)\left(\varepsilon-\varepsilon^{}_{b}\right)}U^{}_{b1}U^{}_{ab}U^{}_{0a}+\frac{J^{}_{0c}J^{}_{cd}J^{}_{d1}}{J^{2}_{cd}-\left(\varepsilon-\varepsilon^{}_{c}\right)\left(\varepsilon-\varepsilon^{}_{d}\right)}U^{}_{d1}U^{}_{cd}U^{}_{0c}\nonumber\\
&\equiv\gamma^{}_{\rm{upper}}U^{}_{\rm{upper}}+\gamma^{}_{\rm{lower}}U^{}_{\rm{lower}}.
\end{eqnarray}
Here, the coefficients $\gamma^{}_{\rm{upper}}$ and
$\gamma^{}_{\rm{lower}}$ are defined as
\begin{eqnarray}
\label{gamma}
\gamma^{}_{\rm{upper}}=\frac{J^{}_{0a}J^{}_{ab}J^{}_{b1}}{J^{2}_{ab}-\left(\varepsilon-\varepsilon^{}_{a}\right)\left(\varepsilon-\varepsilon^{}_{b}\right)},\nonumber\\
\gamma^{}_{\rm{lower}}=\frac{J^{}_{0c}J^{}_{cd}J^{}_{d1}}{J^{2}_{cd}-\left(\varepsilon-\varepsilon^{}_{c}\right)\left(\varepsilon-\varepsilon^{}_{d}\right)},
\end{eqnarray}
and $U^{}_{\rm{lower}}=U^{}_{d1}U^{}_{cd}U^{}_{0c}$,
$U^{}_{\rm{upper}}=U^{}_{b1}U^{}_{ab}U^{}_{0a}$ are the unitary
matrices corresponding to transitions through the lower and upper
paths of the interferometer, respectively. Equations
(\ref{Effective TB}) describe the effective tight-binding
equations for hopping between sites $0$ and $1$ and have the same
form as in the diamond interferometer \cite{AA11,MS13}. All the
details of the interferometer are embodied in the effective
hopping matrix $\bi{W}$ and effective site energies $y^{}_{0}$ and
$y^{}_{1}$.

We next consider the scattering of a wave coming from the left,
i.e.\
\begin{eqnarray}
\label{left wave}
|\psi^{}_{n}\rangle=|\chi^{}_{\rm{in}}\rangle e^{ikna}+r|\chi^{}_{\rm{r}}\rangle e^{-ikna}, \qquad n\leq 0, \nonumber\\
|\psi^{}_{n}\rangle=t|\chi^{}_{\rm{t}}\rangle
e^{ik\left(n-1\right)a}, \qquad n\geq 1,
\end{eqnarray}
where $|\chi^{}_{\rm{in}}\rangle$, $|\chi^{}_{\rm{r}}\rangle$ and
$|\chi^{}_{\rm{t}}\rangle$ are the incoming, reflected and
transmitted normalized spinors, respectively, with the
corresponding reflection and transmission amplitudes $r$ and $t$.
Substituting equations (\ref{left wave}) into (\ref{Effective
TB}), one finds
\begin{eqnarray}
\label{Transmission and Reflection}
t|\chi^{}_{\rm{t}}\rangle\equiv\mathcal{T}|\chi^{}_{\rm{in}}\rangle,
\qquad
r|\chi^{}_{\rm{r}}\rangle\equiv\mathcal{R}|\chi^{}_{\rm{in}}\rangle,
\end{eqnarray}
with the $2\times 2$ transmission and reflection amplitude
matrices
\begin{eqnarray}
\label{Transmission}\mathcal{T}=2ij\sin\left(ka\right)\bi{W}\left(Y\bi{I}-\bi{W}^{\dag}\bi{W}\right)^{-1}, \\
\label{Reflection}\mathcal{R}=-\bi{I}-2ij\sin\left(ka\right)X^{}_{1}\left(Y\bi{I}-\bi{W}^{\dag}\bi{W}\right)^{-1}.
\end{eqnarray}
Here we define
\begin{eqnarray}
\label{X and Y} X^{}_{0,1}=y^{}_{0,1}+je^{-ika}, \quad
Y=X^{}_{0}X^{}_{1}.
\end{eqnarray}
Using equation (\ref{W}), the matrix $\bi{W}^{\dag}\bi{W}$
involved in both $\mathcal{T}$ and $\mathcal{R}$, is found to be
\begin{eqnarray}
\label{WdagW}
\bi{W}^{\dag}\bi{W}=\gamma^{2}_{\rm{upper}}+\gamma^{2}_{\rm{lower}}+\gamma^{}_{\rm{upper}}\gamma^{}_{\rm{lower}}\left(u+u^{\dag}\right),
\end{eqnarray}
with $u=U^{\dag}_{\rm{upper}}U^{}_{\rm{lower}}$ the unitary matrix
representing anticlockwise hopping from site $0$ back to site $0$
around the loop. Equation (\ref{Rotation matrix2}) then yields
$u+u^{\dag}=2\left(\cos\omega\cos\phi+\sin\omega\sin\phi\:\hat{\bomega}\cdot\bsigma\right)$
and equation (\ref{WdagW}) can thus be written as
\begin{eqnarray}
\label{WdagW2} \bi{W}^{\dag}\bi{W}=A+\bi{B}\cdot\bsigma,
\end{eqnarray}
with
\begin{eqnarray}
\label{A and B}
A=\gamma^{2}_{\rm{upper}}+\gamma^{2}_{\rm{lower}}+2\gamma^{}_{\rm{upper}}\gamma^{}_{\rm{lower}}\cos\omega\cos\phi, \nonumber\\
\bi{B}=2\gamma^{}_{\rm{upper}}\gamma^{}_{\rm{lower}}\sin\omega\sin\phi\:\hat{\bi{n}}\equiv
B\hat{\bi{n}}.
\end{eqnarray}
Here, $\hat{\bi{n}}\equiv\hat{\bomega}$ is a real unit vector
along the direction of $\bomega$. As shown in \cite{AA11}, the
spin-dependent transmission of the interferometer is determined by
the eigenvalues of the matrix $\bi{W}^{\dag}\bi{W}$. These
eigenvalues are given by
\begin{eqnarray}
\label{WdagW eigen}
\bi{W}^{\dag}_{}\bi{W}|\pm\hat{\bi{n}}\rangle=\lambda^{}_{\pm}|\pm\hat{\bi{n}}\rangle, \nonumber\\
\lambda^{}_{\pm}=A\pm
B=\gamma^{2}_{\rm{lower}}+\gamma^{2}_{\rm{upper}}+2\gamma^{}_{\rm{lower}}\gamma^{}_{\rm{upper}}\cos\left(\phi\mp\omega\right),
\end{eqnarray}
where $|\pm\hat{\bi{n}}\rangle$ are the eigenstates of the spin
component along the unit vector $\hat{\bi{n}}$, i.e.\
$\hat{\bi{n}}\cdot\bsigma|\pm\hat{\bi{n}}\rangle=\pm|\pm\hat{\bi{n}}\rangle$.
For an incoming spinor $|\pm\hat{\bi{n}}\rangle$ the corresponding
transmission amplitudes $t^{}_{\pm}$ are \cite{AA11}
\begin{eqnarray}
\label{transmission amplitudes}
|t^{}_{\pm}|=\frac{2j\sin\left(ka\right)}{|Y-\lambda^{}_{\pm}|}\sqrt{\lambda^{}_{\pm}},
\end{eqnarray}
and the outgoing electrons are polarized along a different
direction $\pm\hat{\bi{n}}'$, i.e.\ their spinor is
$|\chi^{\rm{out}}_{\pm}\rangle=|\pm\hat{\bi{n}}'\rangle$. Hence,
the transmission amplitude matrix (\ref{Transmission}) can be
rewritten as
\begin{eqnarray}
\label{transmission amplitude matrix}
\mathcal{T}=t^{}_{-}|-\hat{\bi{n}}'\rangle\langle-\hat{\bi{n}}|+t^{}_{+}|\:\hat{\bi{n}}'\rangle\langle\hat{\bi{n}}|.
\end{eqnarray}
Here, $|\pm\hat{\bi{n}}'\rangle$ are the eigenstates of the matrix
$\bi{W}\bi{W}^{\dag}$ \cite{AA11}, namely
\begin{eqnarray}
\label{WWdag eigen}
\bi{W}\bi{W}^{\dag}|\pm\hat{\bi{n}}'\rangle=\lambda^{}_{\pm}|\pm\hat{\bi{n}}'\rangle,
\end{eqnarray}
where the eigenvalues $\lambda^{}_{\pm}$ are given by equations
(\ref{WdagW eigen}). Using equation (\ref{W}), the matrix
$\bi{W}\bi{W}^{\dag}$ is given by
\begin{eqnarray}
\label{WWdag}
\bi{W}\bi{W}^{\dag}=\gamma^{2}_{\rm{upper}}+\gamma^{2}_{\rm{lower}}+\gamma^{}_{\rm{upper}}\gamma^{}_{\rm{lower}}\left(u'+u'^{\dag}\right),
\end{eqnarray}
with $u'=U^{}_{\rm{upper}}U^{\dag}_{\rm{lower}}$ the unitary
matrix representing clockwise hopping from site $1$ back to site
$1$ around the loop.

It is important to emphasize that the spinors
$|\pm\hat{\bi{n}}\rangle$ and $|\pm\hat{\bi{n}}'\rangle$, being
the eigenstates of $u+u^{\dag}$ and $u'+u'^{\dag}$, respectively,
are completely determined by the AB and AC phases and are
independent of the electron energy $\varepsilon$. As we show
below, this implies that spin filtering can be achieved
independent of energy, with the spin polarization direction
controlled solely by the external electric and magnetic fields
(see below). In the next subsection we analyze the general
conditions for spin filtering arising from the transmission
amplitude matrix (\ref{transmission amplitude matrix}) with the
transmission amplitudes (\ref{transmission amplitudes}).
\subsection{General conditions for spin filtering in the double-dot interferometer}
\label{Sec 1B} The spin-polarized current (along $\hat{\bi{n}}'$)
at the output of the interferometer is given by \cite{EO10}
\begin{eqnarray}
\label{spin current}
I=\frac{e}{\hbar}\int^{\infty}_{-\infty}\frac{d\varepsilon}{2\pi}\left[f^{}_{\rm{L}}(\varepsilon)-f^{}_{\rm{R}}(\varepsilon)\right]P^{}_{\hat{\bi{n}}'}(\varepsilon)\Tr\left[\mathcal{T}^{\dag}\mathcal{T}\right],
\end{eqnarray}
where
$f^{}_{\rm{L},\rm{R}}(\varepsilon)=\left[1+e^{\left(\varepsilon-\mu^{}_{\rm{L},\rm{R}}\right)/k^{}_{\rm{B}}T}\right]^{-1}$
is the Fermi distribution in the left (L) or right (R) lead with
the corresponding chemical potential $\mu^{}_{\rm{L}}$ and
$\mu^{}_{\rm{R}}$, $k^{}_{\rm{B}}$ is the Boltzmann constant and
$T$ is the temperature. The spin polarization
$P^{}_{\hat{\bi{n}}'}(\varepsilon)$ along $\hat{\bi{n}}'$ is
defined as
\begin{eqnarray}
\label{spin polarization}
P^{}_{\hat{\bi{n}}'}(\varepsilon)\equiv\frac{\Tr\left[\mathcal{T}^{\dag}\mathcal{T}\bsigma\cdot\hat{\bi{n}}'\right]}{\Tr\left[\mathcal{T}^{\dag}\mathcal{T}\right]}=\frac{|t^{}_{+}|^{2}-|t^{}_{-}|^{2}}{|t^{}_{+}|^{2}+|t^{}_{-}|^{2}},
\end{eqnarray}
where in the last step we used equation (\ref{transmission
amplitude matrix}). For $P^{}_{\hat{\bi{n}}'}(\varepsilon)=\pm 1$
the outgoing electrons with energy $\varepsilon$ are fully
polarized along $\pm\hat{\bi{n}}'$. This occurs if and only if
$|t^{}_{\mp}|=0$, or equivalently $\lambda^{}_{\mp}=0$ [equation
(\ref{transmission amplitudes})]. Without loss of generality, let
us consider the case $\lambda^{}_{-}=0$, in which the outgoing
electrons are polarized along $\hat{\bi{n}}'$. From equation
(\ref{WdagW eigen}) it follows that $\lambda^{}_{\pm}\geq 0$ and
the equality $\lambda^{}_{-}=0$ occurs only if
\begin{eqnarray}
\label{filtering conditions}
\gamma^{}_{\rm{lower}}=\gamma^{}_{\rm{upper}}\equiv\gamma,\nonumber\\
\cos(\phi+\omega)=-1.
\end{eqnarray}
It should be noted that the spin filtering conditions
(\ref{filtering conditions}) are valid for an arbitrary two-path
interferometer with SOI and AB flux. The details of a specific
interferometer enter through the AC phase $\omega$ and the
effective hopping amplitudes $\gamma^{}_{\rm{lower}}$ and
$\gamma^{}_{\rm{upper}}$. The first condition in equations
(\ref{filtering conditions}) can be interpreted as a requirement
for a symmetry relation between the two paths. The second
condition in equations (\ref{filtering conditions}), namely
$\omega=-\phi+\pi$, imposes a relation between the AB flux and the
SOI strength.

To achieve an outgoing spin-polarized beam at finite temperature
or bias voltage, the spin polarization
$P^{}_{\hat{\bi{n}}'}(\varepsilon)$ should be equal to unity in
the relevant energies $\mu^{}_{\rm{R}}-(\rm a\ \rm f \rm e\rm w\
k^{}_{\rm{B}}T)<\varepsilon<\mu^{}_{\rm{L}}+(\rm a\ \rm f \rm e\rm
w\ k^{}_{\rm{B}}T)$ in which electron transport occurs. Therefore,
it is desirable that the spin filtering conditions (\ref{filtering
conditions}) will be satisfied independent of energy. Since the
second condition in equations (\ref{filtering conditions}) is
energy independent, we focus for the moment on the first
condition. From equations (\ref{gamma}), one readily sees that
this condition holds independent of energy if
\begin{eqnarray}
\label{gamma condition}
J^{}_{0a}J^{}_{ab}J^{}_{b1}=J^{}_{0c}J^{}_{cd}J^{}_{d1},\nonumber\\
\varepsilon^{}_{a}+\varepsilon^{}_{b}=\varepsilon^{}_{c}+\varepsilon^{}_{d},\nonumber\\
J^{2}_{ab}-\varepsilon^{}_{a}\varepsilon^{}_{b}=J^{2}_{cd}-\varepsilon^{}_{c}\varepsilon^{}_{d}.
\end{eqnarray}
Compared to the corresponding relations in the diamond
interferometer \cite{AA11,MS13}, equations (\ref{gamma condition})
allow much more freedom for the values of the various parameters
(see below). Therefore, spin filtering can be achieved even in a
very asymmetric interferometer. Furthermore, as argued at the end
of the previous subsection, the spinor of the outgoing electrons,
$|\hat{\bi{n}}'\rangle$, is independent of energy. Hence, the spin
filtering is energy independent provided that equations
(\ref{gamma condition}) hold.

To satisfy equations (\ref{gamma condition}), one can adopt
several approaches. One possibility is to use a single gate
electrode for each branch of the interferometer, so that
$\varepsilon^{}_{a}=\varepsilon^{}_{b}\equiv\varepsilon^{}_{ab}$
and
$\varepsilon^{}_{c}=\varepsilon^{}_{d}\equiv\varepsilon^{}_{cd}$.
The conditions (\ref{gamma condition}) then read
\begin{eqnarray}
\label{gamma condition2}
J^{}_{0a}J^{}_{b1}=J^{}_{0c}J^{}_{d1},\nonumber\\
\varepsilon^{}_{ab}=\varepsilon^{}_{cd},\nonumber\\
J^{}_{ab}=J^{}_{cd}.
\end{eqnarray}
The first condition in equations (\ref{gamma condition2}) can be
satisfied by properly tuning the hopping amplitudes from the leads
to the QDs/QNs, as shown in several experiments
\cite{WFR95,KK02,KY10,YM12}. The second condition can be satisfied
by tuning the gate electrodes. The third condition can be
satisfied by controlling the potential barrier between sites $a$
and $b$ (and/or $c$ and $d$). A further possibility to satisfy
equations (\ref{gamma condition}) is by using two gate electrodes
on each branch of the interferometer \cite{FC07}. Then, by tuning
two site energies, say $\varepsilon^{}_{a}$ and
$\varepsilon^{}_{c}$, one can fulfill the second and the third
conditions in equations (\ref{gamma condition}). The first
condition is again satisfied by tuning one of the hopping
amplitudes from the leads to the QDs/QNs, say $J^{}_{b1}$.

Alternatively, one can work at low temperatures in the
linear-response regime, where all the electrons have the same
energy, equal to the Fermi energy of the leads
$\varepsilon^{}_{\rm{F}}$. The first condition in equations
(\ref{filtering conditions}) should then be satisfied for a single
specific energy $\varepsilon=\varepsilon^{}_{\rm{F}}$. Setting
$\varepsilon^{}_{\rm{F}}=0$ in equations (\ref{gamma}), one has
\begin{eqnarray}
\label{gamma condition3}
\frac{J^{}_{0a}J^{}_{ab}J^{}_{b1}}{J^{2}_{ab}-\varepsilon^{}_{a}\varepsilon^{}_{b}}=\frac{J^{}_{0c}J^{}_{cd}J^{}_{d1}}{J^{2}_{cd}-\varepsilon^{}_{c}\varepsilon^{}_{d}}.
\end{eqnarray}
In this case one has to tune only a single site energy
$\tilde{\varepsilon}=\varepsilon^{}_{a}=\varepsilon^{}_{b}$. In
addition, one should tune the magnetic field or the electric field
perpendicular to the plane in order to satisfy the second
condition in equations (\ref{filtering conditions}).

\begin{figure}[ht]
\centering
\includegraphics[width=0.8\textwidth,height=0.35\textheight]{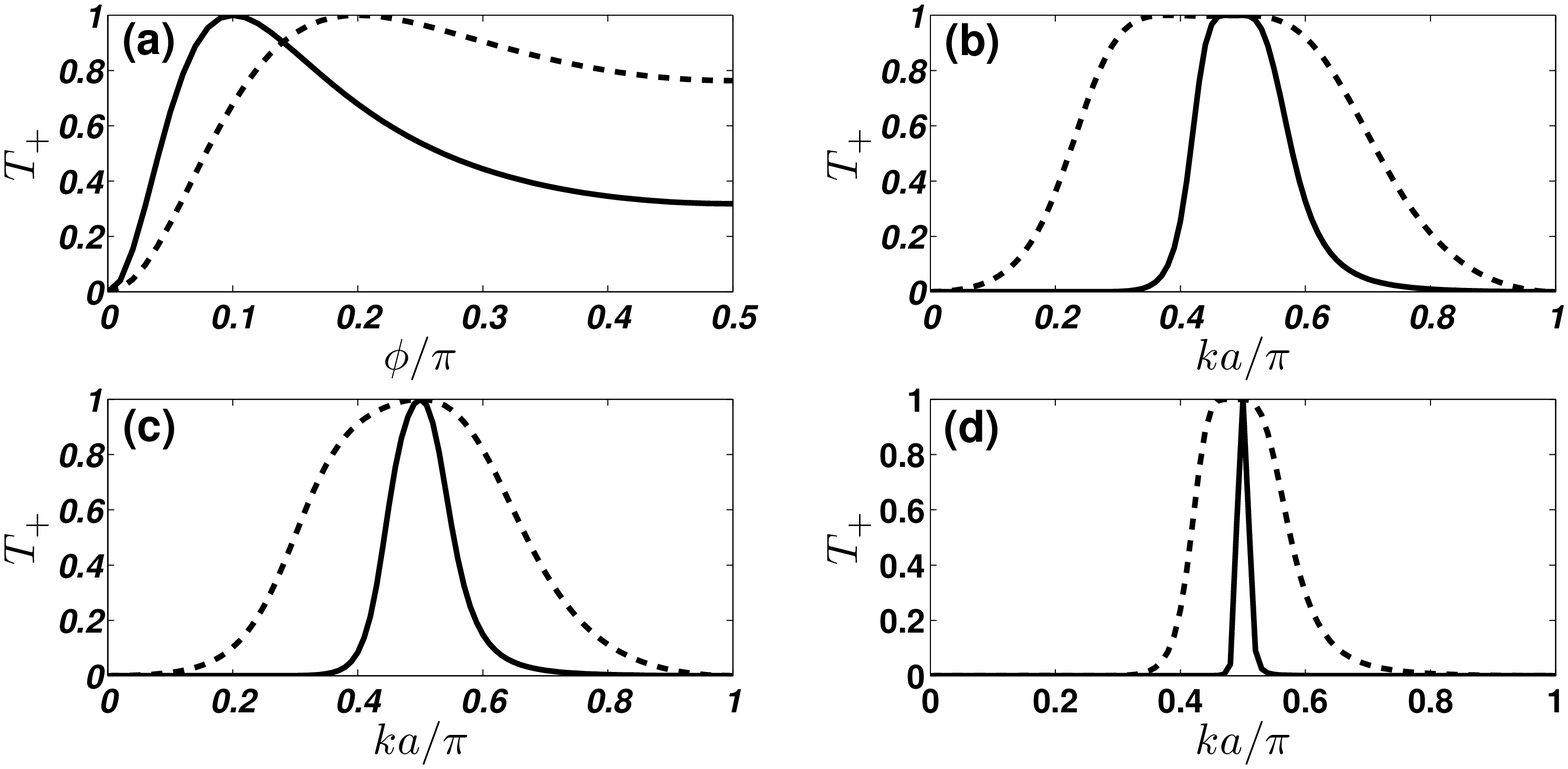}
\caption{\label{fig:transmission} The transmission of the
polarized electrons, $T^{}_{+}(\epsilon,\phi)$ (a) as a function
of the AB flux $\phi$ (in units of $\pi$) for $\epsilon=0$
($ka=\pi/2$) and (b), (c), (d) as a function of $ka$ (in units of
$\pi$) for $\phi=\phi^{}_{0}$. Solid and dashed curves correspond
to maxima of $T^{}_{+}(\epsilon=0,\phi)$ at $\phi^{}_{0}=0.1\pi$
and $\phi^{}_{0}=0.2\pi$, respectively. This is achieved by
choosing
$\gamma^{}_{0}=J^{3}_{1}/J^{2}_{2}=j/\left(2\sin\phi^{}_{0}\right)$,
$\varepsilon^{}_{0}=-\left(J^{2}_{0a}\varepsilon^{}_{b}+J^{2}_{0c}\varepsilon^{}_{d}\right)\gamma^{}_{0}/J^{3}_{1}$
and
$\varepsilon^{}_{1}=-\left(J^{2}_{b1}\varepsilon^{}_{a}+J^{2}_{d1}\varepsilon^{}_{c}\right)\gamma^{}_{0}/J^{3}_{1}$.
The values of the hopping amplitudes and the other site energies
are (b) $J^{}_{0a}=J^{}_{b1}=J^{}_{0c}=J^{}_{d1}=2j$,
$J^{}_{ab}=J^{}_{cd}=8j$ and
$\varepsilon^{}_{a}=\varepsilon^{}_{b}=\varepsilon^{}_{c}=\varepsilon^{}_{d}=\sqrt{J^{2}_{ab}-J^{3}_{1}/\gamma^{}_{0}}$,
(c) $J^{}_{0a}=j$, $J^{}_{b1}=4j$, $J^{}_{0c}=J^{}_{d1}=2j$,
$J^{}_{ab}=J^{}_{cd}=8j$ and
$\varepsilon^{}_{a}=\varepsilon^{}_{b}=\varepsilon^{}_{c}=\varepsilon^{}_{d}=\sqrt{J^{2}_{ab}-J^{3}_{1}/\gamma^{}_{0}}$,
(d) $J^{}_{0a}=J^{}_{b1}=J^{}_{d1}=2j$, $J^{}_{0c}=1.6j$,
$J^{}_{ab}=8j$, $J^{}_{cd}=10j$ and $\varepsilon^{}_{a}=j$.}
\end{figure}

Having fulfilled the spin filtering conditions (\ref{filtering
conditions}), one would like to optimize the transmission
$T^{}_{+}=|t^{}_{+}|^{2}$ of the polarized electrons. Using
equation (\ref{transmission amplitudes}), the transmission has the
form \cite{AA11}
\begin{eqnarray}
\label{Transmission2}
T^{}_{+}=|t^{}_{+}|^{2}=\frac{4j^{2}\sin^{2}\left(ka\right)\lambda^{}_{+}}{P+Q\cos\left(ka\right)+R\cos\left(2ka\right)},
\end{eqnarray}
where
\begin{eqnarray}
\label{Transmission parameters}
P=\left(y^{}_{0}y^{}_{1}-\lambda^{}_{+}\right)^{2}+\left(y^{}_{0}+y^{}_{1}\right)^{2}j^{2}+j^{4},\nonumber\\
Q=2j\left(y^{}_{0}y^{}_{1}-\lambda^{}_{+}+j^{2}\right)\left(y^{}_{0}+y^{}_{1}\right),\nonumber\\
R=2j^{2}\left(y^{}_{0}y^{}_{1}-\lambda^{}_{+}\right).
\end{eqnarray}
The dependence of the transmission on the magnetic flux is only
through $\lambda^{}_{+}$. Substituting $\omega=-\phi+\pi$ and
$\gamma^{}_{\rm{lower}}=\gamma^{}_{\rm{upper}}=\gamma$ into
equations (\ref{WdagW eigen}), one has
$\lambda^{}_{+}=4\gamma^{2}\sin^{2}\phi$. Since one does not
expect the tight-binding model to be valid near the band edges, we
confine ourselves to the center of the band, $\varepsilon=0$ or
$ka=\pi/2$, where the details of the model chosen are not so
important. At the band center we have
$\gamma(\varepsilon=0)\equiv\gamma^{}_{0}=J^{3}_{1}/J^{2}_{2}$
with $J^{3}_{1}\equiv
J^{}_{0a}J^{}_{ab}J^{}_{b1}=J^{}_{0c}J^{}_{cd}J^{}_{d1}$ and
$J^{2}_{2}\equiv
J^{2}_{ab}-\varepsilon^{}_{a}\varepsilon^{}_{b}=J^{2}_{cd}-\varepsilon^{}_{c}\varepsilon^{}_{d}$,
as required by equations (\ref{gamma condition}). The denominator
in equation (\ref{Transmission}) becomes
$P-R=\Big\{\left[\varepsilon^{}_{0}+\left(J^{2}_{0a}\varepsilon^{}_{b}+J^{2}_{0c}\varepsilon^{}_{d}\right)\gamma^{}_{0}/J^{3}_{1}\right]\left[\varepsilon^{}_{1}+\left(J^{2}_{b1}\varepsilon^{}_{a}+J^{2}_{d1}\varepsilon^{}_{c}\right)\gamma^{}_{0}/J^{3}_{1}\right]-\lambda^{}_{+}-j^{2}\Big\}^{2}+j^{2}\left[\varepsilon^{}_{0}+\varepsilon^{}_{1}+\left(J^{2}_{0a}\varepsilon^{}_{b}+J^{2}_{0c}\varepsilon^{}_{d}+J^{2}_{b1}\varepsilon^{}_{a}+J^{2}_{d1}\varepsilon^{}_{c}\right)\gamma^{}_{0}/J^{3}_{1}\right]^{2}$,
which is minimal at
$\varepsilon^{}_{0}=-\left(J^{2}_{0a}\varepsilon^{}_{b}+J^{2}_{0c}\varepsilon^{}_{d}\right)\gamma^{}_{0}/J^{3}_{1}$
and
$\varepsilon^{}_{1}=-\left(J^{2}_{b1}\varepsilon^{}_{a}+J^{2}_{d1}\varepsilon^{}_{c}\right)\gamma^{}_{0}/J^{3}_{1}$.
In this case the transmission is
$T^{}_{+}=4j^{2}\lambda^{}_{+}/(\lambda^{}_{+}+j^{2})^{2}$, and
this has its maximal value of $1$ at $\lambda^{}_{+}=j^{2}$. For a
specific filter one would usually decide around which flux
$\phi^{}_{0}$ one would like to work. We thus optimize the
transmission for a specific flux $\phi=\phi^{}_{0}$. One has a
perfect transmission $T^{}_{+}(\varepsilon=0,\phi=\phi^{}_{0})=1$
at a flux $\phi=\phi^{}_{0}$ if one tunes the parameters so that
$\gamma^{}_{0}=J^{3}_{1}/J^{2}_{2}=j/\left(2\sin\phi^{}_{0}\right)$,
$\varepsilon^{}_{0}=-\left(J^{2}_{0a}\varepsilon^{}_{b}+J^{2}_{0c}\varepsilon^{}_{d}\right)\gamma^{}_{0}/J^{3}_{1}$
and
$\varepsilon^{}_{1}=-\left(J^{2}_{b1}\varepsilon^{}_{a}+J^{2}_{d1}\varepsilon^{}_{c}\right)\gamma^{}_{0}/J^{3}_{1}$.
With these choices, the transmission
$T^{}_{+}(\varepsilon=0,\phi)$ reads
\begin{eqnarray}
\label{Transmission2}
T^{}_{+}(\varepsilon=0,\phi)=\frac{4\sin^{2}\phi\sin^{2}\phi^{}_{0}}{\left(\sin^{2}\phi+\sin^{2}\phi^{}_{0}\right)^{2}}.
\end{eqnarray}
The transmission (\ref{Transmission2}) is plotted in figure
\ref{fig:transmission}(a) as a function of $\phi$ for two values
of $\phi^{}_{0}$. Figure \ref{fig:transmission}(b) shows the
transmission $T^{}_{+}$ versus $ka$ for the flux fixed at
$\phi=\phi^{}_{0}$ and for
$J^{}_{0a}=J^{}_{b1}=J^{}_{0c}=J^{}_{d1}=2j$,
$J^{}_{ab}=J^{}_{cd}=8j$ and
$\varepsilon^{}_{a}=\varepsilon^{}_{b}=\varepsilon^{}_{c}=\varepsilon^{}_{d}=\sqrt{J^{2}_{ab}-J^{3}_{1}/\gamma^{}_{0}}$.
These values correspond to a completely symmetric interferometer
in which the hopping amplitudes and site energies of the lower and
upper branches are identical. The transmission depends smoothly on
energy and remains close to unity in a range around $ka=\pi/2$
which increases with increasing $\phi^{}_{0}$. As expected, the
transmission in this case resembles the transmission of the
symmetric diamond interferometer \cite{AA11}.
Figures \ref{fig:transmission}(c) and \ref{fig:transmission}(d)
are the same as figure \ref{fig:transmission}(b) but for an
asymmetric interferometer. In figure \ref{fig:transmission}(c) we
set $J^{}_{0a}=j$, $J^{}_{b1}=4j$, $J^{}_{0c}=J^{}_{d1}=2j$,
$J^{}_{ab}=J^{}_{cd}=8j$ and
$\varepsilon^{}_{a}=\varepsilon^{}_{b}=\varepsilon^{}_{c}=\varepsilon^{}_{d}=\sqrt{J^{2}_{ab}-J^{3}_{1}/\gamma^{}_{0}}$.
This choice corresponds to an asymmetric interferometer described
by equations (\ref{gamma condition2}), in which the two branches
have the same site energies
$\varepsilon^{}_{a}=\varepsilon^{}_{b}=\varepsilon^{}_{c}=\varepsilon^{}_{d}$
and hopping amplitudes $J^{}_{ab}=J^{}_{cd}$, but different
hopping amplitudes from the leads, namely $J^{}_{0a}\neq
J^{}_{0c}$ and $J^{}_{b1}\neq J^{}_{d1}$. Figure
\ref{fig:transmission}(d) shows the transmission of a completely
asymmetric interferometer for which
$J^{}_{0a}=J^{}_{b1}=J^{}_{d1}=2j$, $J^{}_{0c}=1.6j$,
$J^{}_{ab}=8j$, $J^{}_{cd}=10j$ and
$\varepsilon^{}_{a}\neq\varepsilon^{}_{b}\neq\varepsilon^{}_{c}\neq\varepsilon^{}_{d}$.
We set $\varepsilon^{}_{a}=j$ and then the values of
$\varepsilon^{}_{b}$, $\varepsilon^{}_{c}$ and
$\varepsilon^{}_{b}$ are determined from equations (\ref{gamma
condition}) and from the equation
$\gamma^{}_{0}=J^{3}_{1}/J^{2}_{2}=j/\left(2\sin\phi^{}_{0}\right)$.
A comparison between figures
\ref{fig:transmission}(b)-\ref{fig:transmission}(d) reveals that
the transmission peak at $ka=\pi/2$ gets narrower as the
interferometer becomes more asymmetric. However, comparing the
solid and dashed curves in figures
\ref{fig:transmission}(b)-\ref{fig:transmission}(d), we see that
the narrowing of the transmission peak can be circumvented by
working at a higher flux $\phi=\phi^{}_{0}$.

The results of this subsection thus suggest that perfect spin
filtering, independent of energy, can be accomplished in an
asymmetric interferometer. By tuning the hopping amplitudes, site
energies and AB flux, one can simultaneously satisfy the
conditions (\ref{gamma condition}) and obtain an ideal
transmission of the spin-polarized electrons. The polarization
direction of these spin-polarized outgoing electrons is discussed
in the next subsection.
\subsection{Spin filtering conditions in the presence of Rashba and Dresselhaus SOIs}
\label{Sec 1C} Let us consider now the form of the unitary
matrices $U^{}_{uv}=e^{i\phi^{}_{uv}+i\bi{K}^{}_{uv}\cdot\bsigma}$
in the presence of Rashba and Dresselhaus SOIs. First, consider
the AB phase $\phi^{}_{uv}=-\frac{e}{\hbar}\int^{v}_{u}\bi{A}\cdot
\bi{dr}$. With the gauge $\bi{A}=-By'\hat{\bi{x}}'$ (figure
\ref{fig:square interferometer2}), the AB phases are nonzero only
for the bonds $ab$ and $cd$, with the latter being
$\phi^{}_{ab}=-\phi^{}_{cd}=\phi/2$. To account for the Rashba and
Dresselhaus SOIs, we denote the angle between the $x$ axis and the
crystallographic $(100)$ axis as $\nu$ (figure \ref{fig:square
interferometer2}). With respect to the crystallographic axes, the
unit vectors along the bonds $ab$ and $cd$ are then
$\hat{\bi{g}}^{}_{ab}=\hat{\bi{g}}^{}_{cd}=(\cos\nu,\sin\nu,0)$.
Using equation (\ref{Rotation matrix1}) and denoting
\begin{eqnarray}
\label{Notations}
\alpha^{2}_{uv}=\alpha^{2}_{\rm{R},uv}+\alpha^{2}_{\rm{D},uv},
\quad
\tan\theta^{}_{uv}=\alpha^{}_{\rm{D},uv}/\alpha^{}_{\rm{R},uv}
\quad (uv=ab,cd),
\end{eqnarray}
one ends up with the unitary matrices
\begin{eqnarray}
\label{Rotation Matrices}
U^{}_{ab}=e^{i\phi/2+i\alpha^{}_{ab}\sigma^{}_{ab}}, \quad
U^{}_{cd}=e^{-i\phi/2+i\alpha^{}_{cd}\sigma^{}_{cd}},\nonumber\\
U^{}_{0a}=U^{}_{b1}=U^{}_{0c}=U^{}_{d1}=\bi{I},
\end{eqnarray}
where
$\sigma^{}_{uv}=-\sin\xi^{}_{uv}\sigma^{}_{x}+\cos\psi^{}_{uv}\sigma^{}_{y}$,
with $\xi^{}_{uv}=\theta^{}_{uv}+\nu$ and
$\psi^{}_{uv}=\theta^{}_{uv}-\nu$ ($uv=ab,cd$). Note that
$\sigma^{2}_{uv}=F^{2}_{uv}=1+\sin\left(2\nu\right)\sin\left(2\theta^{}_{uv}\right)$
and therefore
$e^{i\alpha^{}_{uv}\sigma^{}_{uv}}=c^{}_{uv}+is^{}_{uv}\sigma^{}_{uv}$,
with $c^{}_{uv}=\cos\left(\alpha^{}_{uv}F^{}_{uv}\right)$ and
$s^{}_{uv}=\sin\left(\alpha^{}_{uv}F^{}_{uv}\right)/F^{}_{uv}$. To
identify the AC phase $\omega$ and the blocked and transmitted
spin directions, $-\hat{\bi{n}}$ and $\hat{\bi{n}}'$, we calculate
the matrices
$u=U^{\dag}_{\rm{upper}}U^{}_{\rm{lower}}=U^{\dag}_{ab}U^{}_{cd}$
and
$u'=U^{}_{\rm{upper}}U^{\dag}_{\rm{lower}}=U^{}_{ab}U^{\dag}_{cd}$.
Straightforward algebra yields
\begin{eqnarray}
\label{u Matrices}
u=e^{-i\phi}e^{-i\alpha^{}_{ab}\sigma^{}_{ab}}e^{i\alpha^{}_{cd}\sigma^{}_{cd}}=e^{-i\phi}\left(\delta+i\btau\cdot\bsigma\right),\nonumber\\
u'=e^{i\phi}e^{i\alpha^{}_{ab}\sigma^{}_{ab}}e^{-i\alpha^{}_{cd}\sigma^{}_{cd}}=e^{i\phi}\left(\delta+i\btau'\cdot\bsigma\right),
\end{eqnarray}
where
\begin{eqnarray}
\label{u Matrices2}
\delta=c^{}_{ab}c^{}_{cd}+s^{}_{ab}s^{}_{cd}\left(\sin\xi^{}_{ab}\sin\xi^{}_{cd}+\cos\psi^{}_{ab}\cos\psi^{}_{cd}\right),\nonumber\\
\tau^{}_{x}=-\tau'^{}_{x}=s^{}_{ab}c^{}_{cd}\sin\xi^{}_{ab}-s^{}_{cd}c^{}_{ab}\sin\xi^{}_{cd},\nonumber\\
\tau^{}_{y}=-\tau'^{}_{y}=s^{}_{cd}c^{}_{ab}\cos\psi^{}_{cd}-s^{}_{ab}c^{}_{cd}\cos\psi^{}_{ab},\nonumber\\
\tau^{}_{z}=\tau'^{}_{z}=s^{}_{ab}s^{}_{cd}\left(\sin\xi^{}_{cd}\cos\psi^{}_{ab}-\sin\xi^{}_{ab}\cos\psi^{}_{cd}\right),
\end{eqnarray}
and $\delta^{2}+|\btau|^{2}=1$ from unitarity. Comparing equations
(\ref{u Matrices}) with (\ref{WdagW}), (\ref{WdagW eigen}),
(\ref{WWdag eigen}) and (\ref{WWdag}), one derives the following
relations:
\begin{eqnarray}
\label{AC phase and directions} \cos\omega=\delta, \quad
\hat{\bi{n}}=\hat{\btau}, \quad \hat{\bi{n}}'=-\hat{\btau}'.
\end{eqnarray}
Equations (\ref{u Matrices2}) and (\ref{AC phase and directions})
show that the transmitted spin direction $\hat{\bi{n}}'$ differs
from the blocked one $-\hat{\bi{n}}$ in that the components along
the $x$ and $y$ axes are reversed.

Let us examine several special cases of equations (\ref{u
Matrices2}) and (\ref{AC phase and directions}). First, suppose
that one of the branches of the interferometer, say the lower one,
is free of SOI. Substituting $\alpha^{}_{cd}=0$ (and therefore
$c^{}_{cd}=1$, $s^{}_{cd}=0$), equations (\ref{u Matrices2}) take
the form
\begin{eqnarray}
\label{u Matrices3}
\delta=c^{}_{ab},\nonumber\\
\tau^{}_{x}=-\tau'^{}_{x}=s^{}_{ab}\sin\xi^{}_{ab},\nonumber\\
\tau^{}_{y}=-\tau'^{}_{y}=-s^{}_{ab}\cos\psi^{}_{ab},\nonumber\\
\tau^{}_{z}=\tau'^{}_{z}=0.
\end{eqnarray}
Hence, the AC phase in this case is
\begin{eqnarray}
\label{AC phase one branch}
\omega=\alpha^{}_{ab}F^{}_{ab}=\alpha^{}_{\rm{D},ab}\sqrt{1+2\sin\left(2\nu\right)\alpha^{}_{\rm{R},ab}/\alpha^{}_{\rm{D},ab}+\left(\alpha^{}_{\rm{R},ab}/\alpha^{}_{\rm{D},ab}\right)^{2}}.
\end{eqnarray}
Second, if the Rashba mechanism is the dominant SOI, i.e.\
$\alpha^{}_{\rm{R},uv}\gg\alpha^{}_{\rm{D},uv}$, then
$\theta^{}_{uv}\approx 0$ and equations (\ref{u Matrices2}) are
reduced to
\begin{eqnarray}
\label{u Matrices4}
\delta=\cos\left(\alpha^{}_{\rm{R},ab}-\alpha^{}_{\rm{R},cd}\right),\nonumber\\
\tau^{}_{x}=-\tau'^{}_{x}=\sin\left(\alpha^{}_{\rm{R},ab}-\alpha^{}_{\rm{R},cd}\right)\sin\nu,\nonumber\\
\tau^{}_{y}=-\tau'^{}_{y}=-\sin\left(\alpha^{}_{\rm{R},ab}-\alpha^{}_{\rm{R},cd}\right)\cos\nu,\nonumber\\
\tau^{}_{z}=\tau'^{}_{z}=0.
\end{eqnarray}
The AC phase is then simply
$\omega=\alpha^{}_{\rm{R},ab}-\alpha^{}_{\rm{R},cd}$. In the
opposite limit where
$\alpha^{}_{\rm{D},uv}\gg\alpha^{}_{\rm{R},uv}$, one has
$\theta^{}_{uv}\approx\pi/2$ and equations (\ref{u Matrices2})
give
\begin{eqnarray}
\label{u Matrices5}
\delta=\cos\left(\alpha^{}_{\rm{D},ab}-\alpha^{}_{\rm{D},cd}\right),\nonumber\\
\tau^{}_{x}=-\tau'^{}_{x}=\sin\left(\alpha^{}_{\rm{D},ab}-\alpha^{}_{\rm{D},cd}\right)\cos\nu,\nonumber\\
\tau^{}_{y}=-\tau'^{}_{y}=-\sin\left(\alpha^{}_{\rm{D},ab}-\alpha^{}_{\rm{D},cd}\right)\sin\nu,\nonumber\\
\tau^{}_{z}=\tau'^{}_{z}=0.
\end{eqnarray}
Note that in both limits
$\alpha^{}_{\rm{R},uv}\gg\alpha^{}_{\rm{D},uv}$ and
$\alpha^{}_{\rm{D},uv}\gg\alpha^{}_{\rm{R},uv}$, the AC phase is
$\omega=\alpha^{}_{ab}-\alpha^{}_{cd}$. This is not surprising,
since the Rashba and Dresselhaus interactions are related by a
unitary transformation. Furthermore, equations (\ref{u Matrices4})
and (\ref{u Matrices5}) show that in both limits the polarization
of the outgoing electrons is fixed and determined only by the
orientation of the crystal axes. For
$\alpha^{}_{\rm{R},uv}\gg\alpha^{}_{\rm{D},uv}$ the direction of
spin polarization is
$\hat{\bi{n}}'=-\hat{\btau}'=s^{}_{\omega}(\sin\nu,-\cos\nu,0)$
and for $\alpha^{}_{\rm{D},uv}\gg\alpha^{}_{\rm{R},uv}$ the
direction is
$\hat{\bi{n}}'=-\hat{\btau}'=s^{}_{\omega}(\cos\nu,-\sin\nu,0)$,
where $s^{}_{\omega}\equiv\rm{sign}(sin\omega)$. This is different
from the diamond interferometer in which $\hat{\bi{n}}'$ is a
non-trivial function of the SOI strength in both the Rashba and
Dresselhaus limits. The origin of this difference is the geometry
of the two interferometers. The double-dot interferometer consists
of two parallel bonds while the diamond interferometer consists of
four non-parallel bonds. Hence, one can have a fixed spin
polarization at the output of the interferometer provided that
Rashba SOI dominates over the Dresselhaus SOI, or vice versa.
\section{Summary and discussion}
\label{Summary} We have demonstrated that a double-dot
interferometer, made of two parallel QDs/QNs with strong SOIs and
threaded by an AB flux, can serve as a perfect spin filter. As in
the previously suggested diamond interferometer \cite{AA11,MS13},
spin filtering requires two separate conditions. The first one is
the equality of the effective hopping amplitudes for the two
branches of the interferometer
($\gamma^{}_{\rm{lower}}=\gamma^{}_{\rm{upper}}$), while the
second one imposes a relation between the AB and AC phases
($\omega=-\phi+\pi$). These two conditions are necessary for a
complete destructive interference of a specific spin polarization.

The first condition can be regarded as a requirement for global
symmetry between the two branches of the interferometer. If the
temperature or the bias voltage are not very small, complete spin
filtering arises only if one requires the equality
$\gamma^{}_{\rm{lower}}=\gamma^{}_{\rm{upper}}$ to hold
independent of the electron's energy. This imposes several
relations between the various site energies and hopping
amplitudes. In the previously suggested diamond interferometer
\cite{AA11,MS13}, these relations required a perfect symmetry
between the two branches, i.e.\ the global symmetry condition
$\gamma^{}_{\rm{lower}}=\gamma^{}_{\rm{upper}}$ turned into a
local one, requiring for example, the equality of the site
energies of the dots at the corners of the diamond. Here we have
shown that by enlarging the number of site energies and hopping
amplitudes, spin filtering can be achieved in a very asymmetric
interferometer. Needless to say, this is a very positive feature
of the double-dot interferometer for experimental realizations.
Furthermore, we have shown that by tuning the AB flux, the
transmission of the spin-polarized electrons can still be close to
unity in a wide range of energies, even in the asymmetric
interferometer.

The number of interferometer parameters can be enlarged by working
with elongated QDs or QNs. In such nanostructures, one can define
several electrodes and control different parts of the
nanostructure separately \cite{TS09,KY11,FC07,NPS10}. Moreover,
such systems usually have strong SOIs, with the Rashba SOI mostly
being the dominant one. For instance, in InAs nanowires the Rashba
spin-orbit length $\ell^{}_{\rm{SO,R}}=1/k^{}_{\rm{R}}$ was found
to be $\ell^{}_{\rm{SO,R}}\sim 130-200\rm{nm}$ \cite{FC07,HAE05}
which gives $k^{}_{\rm{R}}\sim 5-7.7\cdot 10^{-3}\rm{nm}^{-1}$ [We
remind the reader that $k^{}_{R}$ characterizes the strength of
the Rashba SOI; see equation (\ref{Rashba Hamiltonian})]. Hence, a
nanowire of length $L\sim 200-300\rm{nm}$ would imply
$\alpha^{}_{\rm{R}}=k^{}_{R}L\sim\pi/2$, as required by the
condition $\omega=-\phi+\pi$. For a loop of area $S\approx L^{2}$,
the magnetic field required to create an AB phase $\phi\sim\pi/2$
is $B\sim 10-30\rm{mT}$. The realization of a spin filter using
such systems thus seems feasible.

Based on the observations above, we suggest the following
experiment which can be carried out, for example, using InAs
elongated QDs/QNs \cite{LD12,TS09,KY11,FC07,NPS10}. Since such
systems are usually operated at low temperatures, it is reasonable
to assume linear-response regime. Then at the first stage, one has
to tune a single site energy, e.g.\
$\tilde{\varepsilon}=\varepsilon^{}_{a}=\varepsilon^{}_{b}$, in
order to satisfy equation (\ref{gamma condition3}). The AC phase
$\omega$ is then fixed, and one has to tune the magnetic field to
satisfy the relation $\omega=-\phi+\pi$. Alternatively, one can
apply a fixed magnetic field, and then tune two site energies (say
$\tilde{\varepsilon}^{}_{1}=\varepsilon^{}_{a}=\varepsilon^{}_{b}$
and
$\tilde{\varepsilon}^{}_{2}=\varepsilon^{}_{c}=\varepsilon^{}_{d}$)
to satisfy both the condition $\omega=-\phi+\pi$ and equation
(\ref{gamma condition3}). Either way, at the linear-response
regime spin filtering requires the tuning of only two parameters.
Moreover, we have recently shown that spin filtering can be
achieved even in leaky interferometers \cite{MS13}. Thus the
experiment suggested above can overcome leakage problems, which
can arise in gated QDs/QNs.

How would one verify that the outgoing electrons are indeed fully
spin-polarized? One possible way is by using the so-called spin
blockade effect. One introduces a quantum dot with a strong
Coulomb interaction on or near the outgoing lead \cite{OT09,OK02}.
Starting with no occupation on this dot, and then increasing the
gate voltage on it to capture one electron from the polarized
flow, will block the current due to Pauli's exclusion principle.
This spin blocking was further demonstrated recently, confirming
the spin filtering of a quantum point contact which contains an
SOI \cite{KS12}.
\ack We acknowledge support from the Israel Science Foundation
(ISF).

\end{document}